\documentstyle[floats,aps,psfig]{revtex}
\begin{document}
%\bibliographystyle{unsrt}
%\draft
\title{Resonance Broadening Induced Nonlinear Saturation
of Kinetic Alfven Turbulence in the Interplanetary Plasma}
\author{M.V.~Medvedev 
%\cite{email,mm} 
%{ } and  P.H.~Diamond 
%\cite{pd} }
\thanks{E-mail address: {\tt mmedvedev@ucsd.edu}}
\thanks{ Also: Institute for Nuclear Fusion,
Russian Research Center ``Kurchatov Institute", Moscow 123182, RUSSIA.}
{ } and  P.H.~Diamond 
\thanks{ Also: General Atomics, San Diego, California 92122.}
}
\address{Physics Department, University of California at San Diego,
La Jolla, California 92093-0319.}
%\\[5pt]
%\em Physics Department, University of California at San Diego\\
%\em La Jolla, California 92093-0319.
%\date{\today}
\author{(published: Phys. Letters A {\bf 219} (5-6), August 1996, 293-298)}
\maketitle

\begin{abstract}
The saturation of ion cyclotron Alfv\'en turbulence excited by
beam particles is investigated using resonance broadening theory.
The stochastic scattering which decorrelates particles, includes both random 
acceleration by electric fields 
and a turbulent magnetic mirroring effect.
Turbulent mirroring is shown to yield non-Gaussian corrections 
to the orbits even if the random electric and magnetic fields 
are Gaussian. 
The predicted steady-state turbulence level exhibits a peaked anglular
distribution, with a maximum near $\Theta\simeq 60^\circ$. 
\end{abstract}
%\pacs{96.50Ek, 52.35Qz, 52.35Ra, 52.40Mj}
%\newpage
%\baselineskip18pt
\vskip5ex

It is well known that the interaction of the solar wind plasma with planetary
 magnetospheres
and cometary plasmas results in a high level of wave activity \cite{1.,2.,3.}.
Gyrating solar wind ions are a possible agent for exciting high-frequency
instabilities. It has been suggested \cite{4.,5.} that these magnetic fluctuations 
result from ion cyclotron instability, which occurs when the solar wind 
plasma interacts with a beam of cometary ions. A significant effort has been made to 
understand the turbulence generated by comet-solar wind interactions using 
quasilinear theory [4-6] as well as by simulation studies
\cite{7.,8.,9.,10.,11.,12.}. Ions are 
implanted in the solar wind plasma by melting and photoionization, thus 
forming an unstable `bump-on-tail' like (i.e. shifted Maxwellian) distribution 
in the solar wind frame \cite{13.,14.}. In the far-upstream vicinity of a comet, 
ions form a ring distribution. When the phase velocity of a circularly polarized 
wave coincides with the parallel velocity of particles, i.e. there is an ion 
cyclotron resonance $\omega-k_\|v_\|=\Omega_i$ between the particles 
and the wave, they resonantly pump the wave. Using quasilinear theory,
saturation is predicted to occur via plateau formation in the distribution function. 
This mechanism is rather slow, with a saturation rate of the inverse plateau 
formation time (i.e. a velocity diffusion time). Saturation via resonance 
broadening is much faster, with a time scale comparable to the instability 
growth rate (i.e. since the resonance width is proportional to the fluctuation
amplitude). Thus, even if macroscopic modification of the distribution 
function ({\sl \'a la} plateau formation) occurs, it should be calculated in the 
presence of a finite width, fluctuation-broadened resonance. Moreover, the 
quasilinear theory predicts high level of turbulence at 
$\Theta\simeq 0^\circ$, only ($\Theta$ is the angle between the wave 
propagation direction and the ambient magnetic field), whereas observational
 data (for example, from the comet P/Grigg-Skjellerup \cite{15.,16.}) 
indicates an the average value of $\Theta$ between $50^\circ -60^\circ$ at 
large distances from the comet ($R>400,000~ km.$ for P/GS) and 
$75^\circ -90^\circ$ closer to the comet. The saturation level 
prediction given here, which is 
based on resonance broadening theory, yields a peaked $\Theta$-profile with the 
maximum value at $\Theta_m\simeq 65^\circ$ for P/GS parameters. Note 
that near-normal angles of propagation can also result from excitation of 
left hand polarized waves by normal Doppler resonance \cite{17.} in the 
case when $v_\| \cos{\Theta}<v_A$. For Jupiter we predict
$\Theta_m^{th}\simeq 60^\circ$ and observations from the Ulysses 
spacecraft give $\Theta_m^{exp}\simeq 55^\circ -60^\circ$ \cite{24.}.

We develop a nonlinear kinetic model, so it is natural to anzatz that the 
high-frequency 
turbulent background is composed of a large number of kinetic shear Alfv\'en 
waves which are excited in plasma by the ion cyclotron instability.
The ion-cyclotron resonance becomes $\omega-k_\|v_\|-\Omega_i\sim
k_\|(v_A-v_{beam})-\Omega_i$. We keep $(k_\bot\rho_i)^2$ terms,
however, since they can be significant at large angles $k_\|/k_\bot\ll1$.
An energetic particle immersed in a such turbulent background is subject to 
stochastic scattering which includes both random acceleration by electric 
fields ($\widetilde{\bf E}, \ {\bf v}_\|\times\widetilde{\bf B}_\bot$) as well
as a stochastic analogue of magnetic mirroring  
(${\bf v}_\bot\times\widetilde{\bf B}_\bot$). Random electric fields result 
in the  diffusion of guiding centers in the perpendicular plane.
However, the stochastic mirroring effect produces parallel diffusion in 
velocity space, giving rise to superdiffusion and rapid decorrelation 
(i.e. $\left<x^2\right>\sim\tau^3$) of particle orbit guiding centers along 
the ambient magnetic field. 
Since turbulent mirroring is an intrinsically nonlinear effect, it 
results in non-Gaussian behavior orbit fluctuation statistics even if the 
turbulent electric and magnetic fields obey Gaussian statistics. The random Doppler 
shifts caused by this scattering decorrelate the particles from resonance.
To calculate this stochastic decorrelation, we employ a 
modified stochastic orbit theory \cite{18.,19.} which was developed for
both low-frequency ($\omega\ll\Omega_i$) instabilities \cite{20.,21.} and 
high-frequency ($\omega\simeq\Omega_i$) instabilities \cite{22.}.
As is shown below, the theory yields results which are in very good 
agreement with the experimental data, in
regards amplitude and $k_\bot/k_\|$ dependence.

Starting from the Vlasov equation,
we separate the distribution function into the averaged 
(over fast time and space scales) part $\left< f({\bf x},{\bf v},t)\right>$, 
the test wave response $f_{\bf k}({\bf x},{\bf v},t)$, and 
background responce. We introduce electric and magnetic fields 
of the test wave in terms of scalar and vector potentials as usual, noting 
$\widetilde{ B}_\|=0~ \to~ {\bf A}_\bot=0$. This follows from the fact that 
kinetic shear Alfv\`en waves are incompressible. 
Formally applying the procedure developed in \cite{22.}, we now arrive at 
the lowest order solution of  the Vlasov equation
\begin{eqnarray}
f_{\bf k}&=&{q\over m}\phi_{\bf k}\Biggl\{ {\cal K}_\bot{1\over  v_\bot}
{\partial\over\partial v_\bot} +i\exp(-i{\bf k}\cdot{\bf x})
\sum_{n=-\infty}^{\infty}J_n^2\!\left({k_\bot v_\bot\over\Omega_i}\right)
R(\widehat\omega)
%\nonumber\\& &{ }\cdot
\left[(\omega-k_\|v_\|){\cal K}_\bot
{1\over v_\bot}{\partial\over\partial v_\bot}
+{\cal K}_\|k_\|{\partial\over\partial v_\|}\right]F_M\Biggr\} ,
\label{f(k)} 
\end{eqnarray}
where ${\cal K}_{\|,\bot}$ are the factors:
${\cal K}_\|=\left[1-(1-k_\bot v_\bot/\omega)(1+k_\bot^2\rho_s^2)\right]$,
\ ${\cal K}_\bot=\left[1-{k_\|v_\|/\omega}(1+k_\bot^2\rho_s^2)\right]$, 
and $\rho_s=c_s/\Omega_i$,  $c_s$ is the sound speed,
$\Omega_i=eB_0/m_i c$ is the ion cyclotron frequency. 
Note that in this formula the resonance function $R(\widehat\omega)$ is 
defined by the Fourier transform
\begin{equation}
R(\widehat\omega)=\int_0^\infty d\tau\exp(i{\widehat\omega}\tau)
\langle\exp{[ i{\bf k}\cdot\delta{\bf x}(t-\tau)]}\rangle 
\label{R}
\end{equation}
of the perturbed orbit propagator and asymptotes to the usual resonant 
denominator of the quasilinear theory when fluctuating turbulent forces 
vanish. Here $\delta{\bf x}$ is the deviation of a particle orbit from the 
unperturbed one, i.e. $\delta{\bf x}(t-\tau)={\bf x}'(t-\tau)-{\bf x}(t-\tau)$ 
and $\widehat\omega=\omega-k_\| v_\|-n\Omega_i$.
Using Eq.\ (\ref{f(k)}), we may find (from the dispersion relation) the linear
growth rate of ion cyclotron instability
\begin{eqnarray}
\gamma_0&=&-{n_b \over n_0}{T_i \over T_b}{{{(\omega-\Omega_i)}^2} 
\over{\left|k_\|\right|v_{T_b}}}{\left({\pi \over 2}\right)}^{1/2}
\exp{\left\{-{{{(\omega-\Omega_i)}^2} \over {2k_\|^2{v_{T_b}}^2}}\right\}}
%\nonumber\\& &{ }\cdot
\left[1-{v_D \over v_A}{(1+k^2_\bot\rho_s^2)}^{1/2}-{{{k^2_\bot}
\rho^2_s}\over \Omega_i}(\omega-k_\|v_D)\right] ,
\label{gamma}
\end{eqnarray}
where $n_0$ and $T_i$ are the density and temperature of a bulk plasma,
$n_b$ and $T_b$ are those of the beam, ${\bf v}_D$ is the longitudinal beam 
drift velocity such that $\left((\omega-k_\| v_D-\Omega_i)/k_\| v_{T_b}
\right)^2\ll 1$, and $v_{T_b}$ is the thermal velocity of the beam particles.

To explore the finite amplitude regime, note that the resonance function $R$ 
is expressed via the perturbation of the orbit function of a
test particle, Eq.\ (\ref{R}). Expanding 
$\langle\exp(i{\bf k}\cdot\delta{\bf x})\rangle$ in a
Taylor series, with the  conditions $\Omega_i\tau\gg1$ and 
$\tau\ll\tau_{kinetic}$ ($\tau_{kinetic}$ is the characteristic time of the 
quasilinear diffusion), we have
\begin{eqnarray}
\langle\exp(i{\bf k}\cdot\delta{\bf x})\rangle&=&\exp\bigl\{\langle i{\bf k}
\cdot\delta{\bf x}\rangle-{1\over 2!}\langle 
({\bf k}\cdot\Delta\delta{\bf x})^2 \rangle+\dots\bigr\} 
\nonumber\\
&=&\exp\{i\delta\omega\tau-k^2_\bot D_\bot\tau-k^2_\| 
D_{v_\|}\tau^3+\dots\} , 
\label{taylor}
\end{eqnarray}
where $\Delta\delta{\bf x}=\delta{\bf x}-\langle\delta{\bf x}\rangle$,
$\Delta\omega_\bot=k^2_\bot D_\bot$,
$D_\bot$ is the perpendicular diffusion coefficient and $D_{v_\|}$ is the
diffusion coefficient of the longitudinal velocity in velocity space.  The cubic in
$\tau$ term gives rize to a $\Gamma$-function in the real-space 
representarion of the resonance function. For the intermediate time-scale 
$\Omega_i^{-1}\ll\tau\ll\tau_{kinetic}$ as above, one can simplify the result 
by assuming $\tau^3 k^2_\| D_{v_\|}\to\tau (k_\|^2 D_{v_\|})^{1/3} $. 
The resonance frequency shift $\delta\omega$ is proportional to spatial 
derivatives of the guiding center distribution, and vanishes for a Maxwellian 
\cite{22.}. Then, using the approximations mentioned above, the resonance
function  becomes
\begin{equation}
R(\widehat\omega)={i\over\widehat\omega+i\Delta\omega_\bot
+i\Delta\omega_\|} ,
\label{R2}
\end{equation}
where $\Delta\omega_\bot=k^2_\bot D_\bot$ and 
$\Delta\omega_\|=(k_\|^2 D_{v_\|})^{1/3}$.

The system of equations describing the perturbed motion of a test particle 
cannot be solved exactly because of existence of the nonlinear coupling term
${\bf v}\times{\bf B}$. We use a perturbation method, assuming small 
perturbations. Up to the second order in perturbation, we have
\begin{eqnarray}
& &{\bf x}(t+\tau)={\bf x}(t)
+\widehat{\bf e}_\bot\left[{v_\bot\over 2}\exp{[i\Delta\psi]}
{{\exp{(i\Omega_i\tau)}-1}\over {i\Omega_i}}+c.c.\right]
+\widehat{\bf e}_\|v_\|\tau
\nonumber\\
& &{ }+\widehat{\bf e}_\bot\left[{1 \over 2}\int_0^{\tau}ds{q \over m}
\Bigl(\widetilde{\cal E}^+(t+s)+\widetilde{\cal F}^+(t+s)\Bigr)\exp{(-i\psi)}
{{\exp{[i\Omega_i(\tau-s)]}-1} \over {i\Omega_i}}+c.c.\right]
\nonumber\\
& &{ }+\widehat{\bf e}_\|\int_0^{\tau}ds(\tau-s){q \over m}\Bigl(
\widetilde{\cal E}_\|(t+s)+\widetilde{\cal F}_\|(t+s)\Bigr) ,
\label{x(t)}
\end{eqnarray}
where $\Delta\psi$ is the angle between ${\bf k}_\bot$
and ${\bf v}_\bot$ and
\begin{equation}
\begin{array}{rl}
&\displaystyle{
\widetilde{\cal E}^\pm_\bot(t)
=\widetilde{E}^\pm_\bot(t)\mp i{v_{\|0} \over c}\widetilde{B}^\pm_\bot(t), 
\quad\widetilde{\cal E}_\|(t)=\widetilde{E}_\|(t)+{i \over{2c}}
\Bigl(v^+_{\bot0}
\widetilde{B}^-_\bot(t)-v^-_{\bot0}\widetilde{B}^+_\bot(t)\Bigr) ,}\\[1.0em]
%\nonumber\\
&\displaystyle{
\widetilde{\cal F}^\pm_\bot(t)=\mp i{\widetilde{v}_\| \over c}
\widetilde{B}^\pm_\bot(t),
\quad\widetilde{\cal F}_\|(t)
={i\over{2c}}\Bigl(\widetilde{v}_\bot^+\widetilde{B}^-_\bot(t)-
\widetilde{v}^-_\bot\widetilde{B}^+_\bot(t)\Bigr) ,}\\[1.0em]
%\nonumber\\
&\displaystyle{
\widetilde{v}_\|(t)={q \over m}\int_0^t\widetilde{E}_\|(t')dt',
\quad\widetilde{v}^\pm_\bot(t)
=e^{-i\Omega_i t}\int_0^te^{i\Omega_i t'}{q \over m}
\Bigl[\widetilde{E}^\pm_\bot(t')\mp i{v_{\|0} \over c}
\widetilde{B}^\pm_\bot(t')\Bigr]dt' .}
\end{array}
\end{equation}
Here $\widehat{\bf e}_\bot$ is the unit vector in the plane perpendicular to 
the ambient magnetic field parallel to ${\bf v}_\bot$,
$\widehat{\bf e}_\|$ is the unit vector collinear to the ambient magnetic field,
$v^\pm_\bot=v_x\pm iv_y$ and $v_{\bot 0}^\pm, v_{\| 0}$ are the zeroth 
order velocities.
From here we see that even if the $\widetilde{\cal E}\propto\phi_{\bf k}$ 
field is
Gaussian, $\widetilde{\cal F}$ is not always Gaussian, but rather given by a 
distribution proportional to a Gaussian multiplied by an error function. This 
means that  the third and other moments in the expansion (\ref{taylor}) are
non-zero. The error function is an increasing function of its argument. Thus,
corrections appear on the tail of the Gaussian distribution,
 i.e. at sufficiently large times such that $\tau\sim\tau_{kinetic}$. This effect 
implies temporal intermittancy for strongly mirrored particles. On 
intermediate time-scales $\Omega_i^{-1}\ll\tau\ll\tau_{kinetic}$, this 
non-Gaussian correction is small even when $\delta B/B\sim1$.
To capture the simple physical effect of resonance  broadening we can drop 
the non-Gaussian terms $\widetilde{\cal F}$ as small 
perturbations. After straightforward calculation using 
Eq.\ (\ref{taylor},\ref{x(t)}) and omitting rapidly oscillating terms in 
$\tau$, we obtain
\begin{eqnarray}
\Delta\omega_\bot&=&{{{k^2_\bot} v^2_A}\over{B_0^2(1+{k^2_\bot}
\rho_s^2)}}
\sum_{{\bf k'}}\sum^{\infty}_{n=-\infty}{\left|B_{\bot{\bf k'}}\right|}^2
\left\{1+{v_\| \over {v_A}}{(1+{{k'}^2_\bot}\rho_s^2)}^{1/2}\right\}
\nonumber\\& &{ }\cdot
{1 \over 4}\left[J_{n-1}^2(s)+2J_n^2(s)+J_{n+1}^2(s)\right]
R(\omega-k_\|v_\|-n\Omega_i) 
\label{om-perp}
\end{eqnarray}
and similarly
\begin{eqnarray}
\Delta\omega_\|&=&\Biggl\{{\Omega_i^2\over B_0^2}{{k^4_\|}\over
{k^2_\bot}}{v_A^2\over (1+{k^2_\bot}\rho_s^2)}\sum_{\bf k'}
\sum_{n=-\infty}^{\infty}{\left|B_{\bot_{\bf k'}}\right|}^2 \biggl(J_n^2(s)
\nonumber\\& &{ }
+{1 \over 4}{{{k'}^2_\bot} v_\bot^2\over {{k'}^2_\|} v_A^2}
\left[J^2_{n-1}(s)-J^2_{n+1}(s)\right]R(\omega-k_\|v_\|-n\Omega_i)
\biggr)
\Biggr\}^{1/3} . 
\label{om-par}
\end{eqnarray}
The sums in these equations run over the full range of the background 
turbulence spectrum $-\infty<k'<\infty$. Here `unprimed' wave vectors refer 
to the test wave. Eqs.\ (\ref{om-perp},\ref{om-par}) describe the general 
case of resonance broadening due to the effects of scattering of a test 
particle by random fields. They can be simplified by 
assuming a narrow-band spectrum 
\cite{22.}.   As it will be shown later from estimating the spectrum at 
saturation, the narrow spectrum approximation is valid for this case. 
Then $\sum_{\bf k}\left|B_{\bot_{\bf k}}\right|^2/{B_0}^2$ can be
replaced by $(\delta B/B)^2$. As we consider the ion cyclotron resonance with 
ions we may set $n=1$. $n\not=1$ terms give only a small correction. 

\begin{figure}
\psfig{file=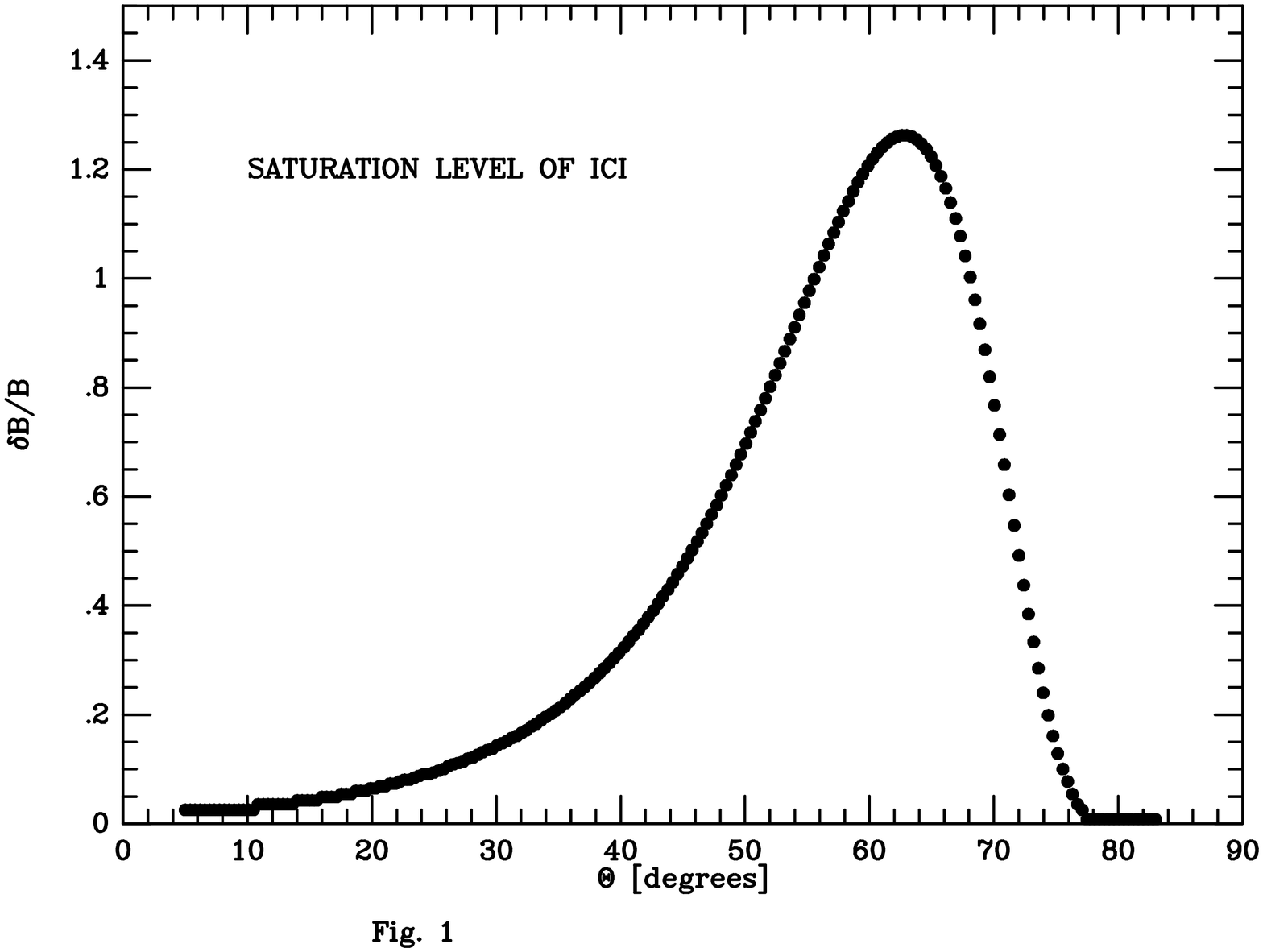,width=4in,angle=-90}
\vskip-42.5ex\hspace*{57ex}
\psfig{file=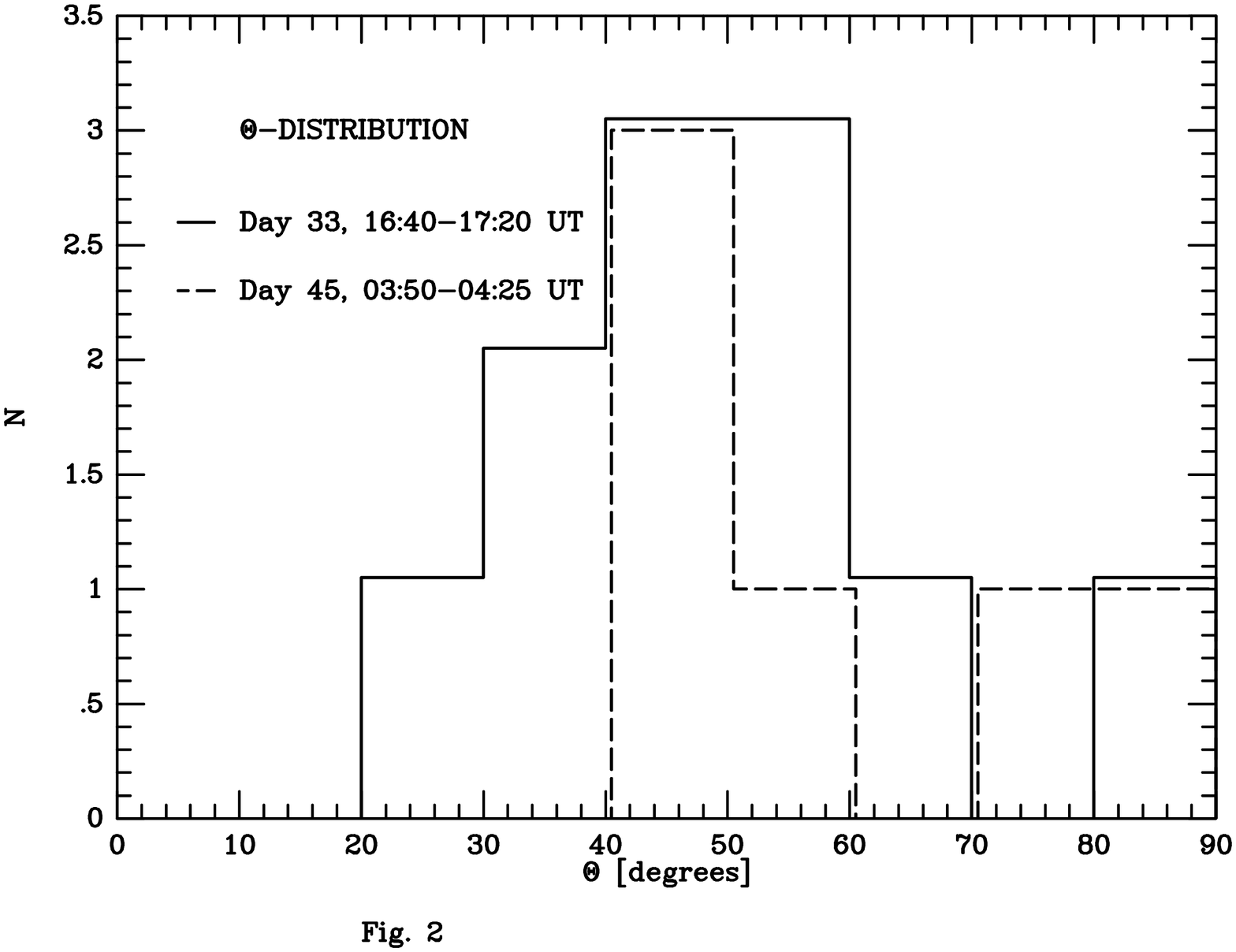,width=4in,angle=-90}
\caption{The angular distribution of the saturated Alfv\'en turbulence 
fluctuation level.}
\label{fig1}
\end{figure}

The broadening $\Delta\omega({\bf v})$ should be averaged over the 
velocity distribution function \cite{22.}. Saturation occurs when the resonance 
broadening becomes comparable to the linear growth rate $\gamma_0$, 
i.e. when the total nonlinear growth rate vanishes,
$\gamma_{NL}=\gamma_0-\overline{\Delta\omega}=0$.
This saturation mechanism does not rely upon quasilinear plateau
formation and thus is applicable to instances where the unstable distribution is 
maintained externally, such as in the case of comet-solar wind interaction.
For the ion cyclotron instability, we define the average broadening
$\overline{\Delta\omega}=\Gamma_1^{-1}(s_i)\int {d{\bf v}
F_M({\bf v})J_1^2
\left({{k_\bot v_\bot}/{\Omega_i}}\right)\Delta\omega({\bf v})}$,     
where $s_i=k_\bot v_{T_i}/\Omega_i$ and 
$\Gamma_n(s_i)=\int d{\bf v}F_M ({\bf v})J_n^2 (k_\bot v_\bot/\Omega_i)$,
$n=0,1,\dots$.
Upon averaging of 
Eqs.\ (\ref{om-perp},\ref{om-par})
with $R(\widehat\omega)$ written for the unperturbed case, we can 
calculate the nonlinear saturation level. We use a perturbation analysis to
explore those regions where the turbulence level is (sufficiently) 
small, i.e. at angles of propagation close to $0^{\circ}$ and $90^{\circ}$ 
(see below). For the case of the perpendicular 
diffusion (i.e. $\gamma_0-\Delta\omega_\bot=0$), we obtain
\begin{equation}
\biggl({{\delta B} \over B}\biggr)^2_{NL_\bot}
={{{(\omega-k_\|v_D-\Omega_i)}^2
(1+{k^2_\bot}\rho^2_s)} \over {{k^2_\bot}v_A^2\left(1+\displaystyle{v_D 
\over v_A}{(1+{k^2_\bot}\rho_s^2)}^{1/2}\right)F_1(s_i)}} , 
\label{B-perp}
\end{equation}
where $v^2_A=B_0^2/4\pi\rho$ is the Alfv\'en speed, $F_1(s_i)
=\Gamma_1^{-1}(s_i)\int 0.5 J_1^2 \left(J^2_0+2J^2_1+J^2_2\right) 
F_M({\bf v})d{\bf v}$. The saturation level is roughly proportional to 
$k^{-2}_\bot\sim\sin^{-2}\Theta$. Thus this case corresponds to the 
saturation at large angles. By contrast, if there is only parallel diffusion
(i.e. $\gamma_0-\Delta\omega_\|=0$), we have
\begin{equation}
\biggl({\delta B \over B}\biggr)^2_{NL_\|}
={\gamma_0^2 \over \Omega_i^2}
{(\omega-k_\|v_D-\Omega_i)^2(1+{k^2_\bot}\rho^2_s){k^2_\bot} \over
{{k^4_\|}v_A^2\left[F_2(s_i)+2\displaystyle{{k^2_\bot}v_{T_i}^2\over 
{k^2_\|}v_A^2} F_3(s_i)\right]}} ,
\label{B-par}
\end{equation}
where
$F_2(s_i)=\Gamma^{-1}(s_i)\int J_1^4 F_M({\bf v})d{\bf v}$,\,
$F_3(s_i)=\Gamma^{-1}(s_i)\int 0.25 J_1^2 \left[J^2_0-J^2_2\right]
F_M({\bf v})v_\bot d{\bf v}$. The fluctuation level $\delta B/B$ in this case 
is proportional to 
$k^{-2}_\|\sim\cos^{-2}\Theta$. Thus parallel diffusion suppress turbulence 
at small angles. In the derivation of this equation we made the  approximation 
$ \overline{f^3}\approx \overline {f}^3$, but indeed they differ only by a 
factor close to unity. The $\Delta\omega_\|$ effect is dominant at small 
angles of wave propagation, whereas the $\Delta\omega_\bot$ effect is 
more important at  near-perpendicular angles. For the case where 
$\Delta\omega=\Delta\omega_\bot  +\Delta\omega_\|$ it is
necessary to solve  the third order algebraic equation of  the type
$\gamma_0=c_1{\delta B}^2+c_2{\delta B}^{2/3}$. 
Fig.\ \ref{fig1} represent the solution of this equation for typical comet and 
solar wind parameters: $n=10~cm^{-3}, n_b=1.7~cm^{-3}, T_e=T_i=10~eV, 
T_b=3~eV, v_{T_i}=3\cdot 10^7~cm/s, v_A=0.2v_{T_i}, v_D=2v_A, 
\Omega_i=10~s^{-1}, k=10^{-5} cm^{-1}$. 

\begin{figure}
\caption{The angular distribution of the MHD activity level (in arbitrary 
units) near Jupiter obtained from Ulysses data.}
\label{fig2}
\end{figure}
Fig.~1 shows the anisotropy of magnetic fluctuation level in
propagation angle $\Theta~(\tan\Theta=k_\bot/k_\|)$. 
 One can see that at small angles $(\Theta\simeq 0^\circ)$ with respect to the
ambient magnetic field, very small background turbulence levels saturate 
the ion  cyclotron instability. The same is true of the near-perpendicular case,
 $\Theta\sim 90^\circ$. There the turbulence is strongly self-suppressed. 
This is  not the case for  $\Theta\simeq 60^\circ$.  Here, the amplitude of 
the background Alfv\'en waves must  be large for resonance
 broadening  to self-consistently suppress the ion cyclotron instability.
The main energy content of the observable  MHD activity, therefore, is 
associated with Alfv\'en waves propagating at large angles relative 
to the averaged magnetic field. The peak is strongly localized at  angles near
$\Theta _m\simeq 60^\circ$. The value $\Theta_m$ of the maximum 
turbulent fluctuation is remarkably insensitive to  plasma parameters such 
as bulk plasma and ion beam temperatures, density and drift velocity of the 
beam, etc. For all reasonable parameters, $\Theta_m$ remains approximately
 60 degrees. The magnitude of the maximum fluctuation level, on
the contrary, depends sensitively on the beam and plasma parameters. 
Fig.\ \ref{fig2} is a plot of experimental data of  MHD activity near Jupiter, 
obtained by the spacecraft Ulysses \cite{24.}. Here the level of fluctuations 
(in arbitrary units), measured for different angles, is shown. 
It is apparent that the dominant waves are those
propagating  at angles in the interval  $50^\circ<\Theta<70^\circ$. 
There is remarkable agreement of the angular distribution obtained from 
experiments \cite{15.,16.,24.} with our very simple theory.
Quasilinear theory predicts a different result \cite{5.,17.}, which is 
(approximately) $(m_in_i/m_pn_p)^{2/3}(v_D/v_A)^{1/3}$. The
$\Theta$-dependence predicted by quasilinear theory is not in agreement 
with experimental data, although it gives a reasonable prediction of 
the magnitude, i.e.  $\delta B/B\sim 1$.

We emphasize the fact that the peaked angular distribution is the result 
of both perpendicular and parallel diffusion. The perpendicular diffusion 
(the diffusion of guiding centers) is a result of the magnetic field line flutter 
and is responsible for the resonance broadening at larger angles, 
$\Theta\sim 90^\circ$. The parallel diffusion (the randomized step-size of
a particle helical trajectory) is due to the random mirroring force. 
It is responsible for 
the instability suppression at small angles of wave propagation, i.e.  for 
$\Theta$ near $0^\circ$. It is important to note that the parallel diffusion 
in velocity space results in a $\left<x^2\right>\sim\tau^3$ type decorrelation 
process in real space. However, for intermediate time-scales 
$\Omega_i^{-1}\ll\tau\ll\tau_{kinetic}$, it can be modelled by a standard 
diffusion process. The magnetic mirroring gives rise to non-Gaussian 
dynamics of a particle immersed into turbulent background, even if this 
background is Gaussian. This describes temporal intermittancy in a system 
with strongly mirrored particles ($\delta B/B\sim 1$).

We are very grateful to S.K.~Ride for discussions and for unpublished data 
from the Ulysses spacecraft. We also thank V.D.~Sapiro, V.I.~Shevchenko 
and B.T.~Tsurutani for useful discussions. This research was supported 
by U.S. Department of Energy, Contract No. DE-FG03-88ER53275 and 
NASA Grant No. UT-A:NAGW-2418.

%\bibliography{ }
%
\end{document}